\documentclass[amsmath,amssymb,pra,aps,showpacs,superscriptaddress,twocolumn,10pt]{revtex4-1}
\usepackage{amsmath,amsfonts,amssymb,amsthm,graphics,graphicx,epsfig,bbm}
\usepackage[colorlinks=true,citecolor=blue,linkcolor=red,urlcolor=blue]{hyperref}
\usepackage[usenames]{color}
\usepackage{graphicx}
\usepackage{subfigure}
\usepackage{amsmath}
\usepackage{epsfig}
\usepackage{dcolumn}
\usepackage{bm}
\usepackage{color}
\usepackage{epstopdf}
\usepackage{amssymb}
\usepackage{amstext}
\usepackage{latexsym}
\usepackage{hyperref}
\usepackage{amsfonts}
\usepackage{psfrag}
\usepackage{xcolor}
\usepackage[normalem]{ulem}
\usepackage{dsfont}
\usepackage{txfonts}
\usepackage{overpic}

\newcommand{\ket}[1]{| #1 \rangle}
\newcommand{\bra}[1]{\langle #1 |}

\newcommand{\ew}[1]{\langle #1 \rangle}
\newcommand{\beq}{\begin{eqnarray}}
\newcommand{\eeq}{\end{eqnarray}}

\begin{document}

%%%%%%%%%%%%%%%%%%%%%%%%%%%%%%%%%
\title{Ergodic-localized junctions in periodically-driven systems} 
%%%%%%%%%%%%%%%%%%%%%%%%%%%%%%%%%
\author{V. M. Bastidas}
 \email{victor.bastidas@lab.ntt.co.jp}
  \affiliation{NTT Basic Research Laboratories \& Research Center for Theoretical Quantum Physics,  3-1 Morinosato-Wakamiya, Atsugi, Kanagawa, 243-0198, Japan} 
  \author{B. Renoust}
   \affiliation{Osaka University, Institute for Datability Science, 2-8 Yamadaoka, Suita, Osaka Prefecture 565-0871, Japan} 
    \affiliation{National Institute of Informatics, 2-1-2 Hitotsubashi, Chiyoda-ku, Tokyo 101-8430, Japan}
    \affiliation{Japanese-French Laboratory for Informatics, CNRS UMI 3527, 2-1-2 Hitotsubashi, Chiyoda-ku, Tokyo 101-8430, Japan}
 \author{Kae Nemoto}
  \affiliation{National Institute of Informatics, 2-1-2 Hitotsubashi, Chiyoda-ku, Tokyo 101-8430, Japan}
  \affiliation{Japanese-French Laboratory for Informatics, CNRS UMI 3527, 2-1-2 Hitotsubashi, Chiyoda-ku, Tokyo 101-8430, Japan} 
 \author{W. J. Munro}
 \affiliation{NTT Basic Research Laboratories \& Research Center for Theoretical Quantum Physics,  3-1 Morinosato-Wakamiya, Atsugi, Kanagawa, 243-0198, Japan} 
  \affiliation{National Institute of Informatics, 2-1-2 Hitotsubashi, Chiyoda-ku, Tokyo 101-8430, Japan}

\date{\today}

%%%%%%%%%%%%%%%%%%%%%%%%%%%%%%%%%
\begin{abstract}
%%%%%%%%%%%%%%%%%%%%%%%%%%%%%%%%%
Quantum phases of matter have many relevant applications in quantum computation and quantum information processing. Current experimental feasibilities in diverse platforms allow us to couple two or more subsystems in different phases. In this letter, we investigate the situation where one couples two domains of a periodically-driven lattice of interacting particles, where one of them is ergodic while the other is fully localized. By combining tools of both graph and Floquet theory, we show that the localized domain remains stable for strong disorder, but as this disorder decreases the localized domain becomes ergodic. 
%%%%%%%%%%%%%%%%%%%%%%%%%%%%%%%%%
\end{abstract}
%%%%%%%%%%%%%%%%%%%%%%%%%%%%%%%%%

\maketitle

%%%%%%%%%%%%%%%%%%%%%%%%%%%%%%%%%%%%%%%%%%%%
%\textit{Introduction:\textemdash}
%%%%%%%%%%%%%%%%%%%%%%%%%%%%%%%%%%%%%%%%%%%%
One of the most intriguing aspects of physics is the nontrivial collective behavior of matter at low temperatures~\cite{Shahar1997}. In contrast to classical phase transitions, quantum fluctuations can induce changes between different quantum phases of matter at temperatures close to the absolute zero~\cite{Shahar1997}. For instance, quantum interference is responsible for the metal-to-insulator transition in the Anderson model of electrons moving in disordered potential~\cite{anderson58}. Since its discovery, Anderson localization has been a paradigmatic phenomenon in condensed matter physics~\cite{kramer93,Mirlin2008} 
and it has had dramatic consequences. In one and two dimensions, uncorrelated disorder leads to a massive localization of all the energy states and the system behaves always as an insulator in the thermodynamic limit~\cite{anderson58,kramer93,Mirlin2008}. However, there are well-known exceptions to this behavior, including  when the disorder has certain correlations~\cite{aubry80,Sokolov1999,Krokhin1999,Miniatura2009,Vignolo2015}, or if there is long-range hopping along the lattice~\cite{anderson58,Dominguez-Adame2003,Borgonovi2016}.
Recently there has been an enormous amount of interest on the effect of interactions on localization properties of the states. The interplay between disorder and interactions can be exploited to avoid thermalization in many-body systems, which is referred to as many-body localization (MBL)~\cite{Altshuler2006, nandkishore15,Vosk2015,khemani17}. This phenomenon is closely related to the Anderson model in random graphs~\cite{Scardicchio2014,Skvortsov2016,Kravtsov2016,Slanina2017,Lemarie2017,Mirlin2017} and has been observed  in trap ions~\cite{Monroe2016}, cold atoms~\cite{Bloch2015,Gross2016}, and superconducting qubits~\cite{roushan17,Fan2018}.

In recent years, there has been an increasing interest in the exploration of localization properties of many-body systems under the effect of an external driving~\cite{Polkovnikov2013,Moessner2015,Abanin2015,Moessner2016,Brandes2016,Refael2018}. The drive can produce unexpected effects~\cite{Polkovnikov2015,Eckardt2017} and states of matter that are absent in undriven systems such as discrete time crystals arise~\cite{else16,yao17,Zakrzewski2018}. In addition, an external drive can suppress tunneling in a coherent way, which is referred to as coherent destruction of tunneling~\cite{Hanggi1991,Haenggi1998}. This can be used to generate a nonequilibrium version of the Mott-insulator transition~\cite{Holthaus2005}, which has observed in driven optical lattices\cite{Arimondo2007}. Moreover, in contrast to undriven models, many-body systems can absorb energy from the external drive and so heat up to infinite temperature~\cite{Moessner2014,Rigol2014}. This phenomenon is accompanied by the divergence of celebrated high-frequency expansions~\cite{Anisimovas2015,Saito2016}. Hence the stroboscopic dynamics cannot be described using a local Hamiltonian~\cite{Moessner2014,Rigol2014}.

In this letter, we investigate the interplay between driving and disorder in a one-dimensional lattice of interacting bosons that can be realized currently in diverse platforms~\cite{Bloch2015,Gross2016,Monroe2016,roushan17,Martinis2018}. We consider a lattice of interacting bosons that is divided into two domains: half of the lattice is disordered while the second half is driven. The total system can be thought as an ergodic-localized (EL) junction where the drive leads to ergodicity in one domain, while disorder induces localization in the other one. In the context of undriven systems, this situation resembles the so-called many-body localization proximity effect, where a many-body-localized subsystem is coupled to a thermalized one~\cite{Nandkishore2015,Nandkishore2018}. While in some cases the thermalized system becomes localized~\cite{Nandkishore2015}, there is numerical evidence of thermalization of the whole system~\cite{Bauer2017}. Recently, the stability of the localized phase has been the focus of active theoretical~\cite{DeRoeckHuveneer2017,HuveneersStab2017,Chandran2017,DeRoeck2017} and experimental~\cite{Schneider2016,Schneider2017,Gross2018,Bloch2018} research. The theoretical description of localization properties of a EL junction changes dramatically in the context of periodically-driven systems and there are nontrivial effects that do not appear in the undriven case~\cite{Polkovnikov2015,Brandes2016,Eckardt2017}. In this work,  we show that if one looks at the total system stroboscopically, the dynamics is generated by a highly non-local Hamiltonian. We provide a geometrical interpretation of the total system by using graph theory tools. When the disorder is weak, there is a proximity effect where the localized domain becomes unstable. The stability of the latter increases by increasing the disorder. The EL junction can be visualized as a graph with two clusters: the localized domain is a cluster with low connectivity sites, whereas the ergodic domain is highly connected. The proximity effect is related to an increasing connectivity between the two clusters.   We show that the latter is related to the participation ratio, that is, sites with high connectivity in the graph have also a high participation ratio.

%%%%%%%%%%%%%%%%%%%%%%%%%%%%%%%%%%%%%%%%%%%%
%\textit{The Hamiltonian of the total system:\textemdash}
%%%%%%%%%%%%%%%%%%%%%%%%%%%%%%%%%%%%%%%%%%%%
Motivated by recent experiments~\cite{Bloch2015,Gross2016,roushan17,Martinis2018}, in this work we consider a system of $N$ interacting bosons in a one-dimensional lattice with $L$ sites given by the Hamiltonian 
%%%%
\begin{align}
\label{eq:1DBosonicGmonArray}
\hat{H}(t)=\hbar\sum^{L}_{j=1}\left[h_{j} \hat{n}_{j}+\frac{U}{2}\hat{n}_{j}(\hat{n}_{j}-1)\right]+\hbar\sum^{L-1}_{j=1}g_j(t)(\hat{a}^{\dagger}_{j}\hat{a}_{j+1}+\text{H.c})
 \ .
  \end{align}
%%%%
Here $\hat{a}_{j}$ and $\hat{a}^{\dagger}_{j}$ are the bosonic anhililation and creation operators, $\hat{n}_{j}=\hat{a}^{\dagger}_{j}\hat{a}_{j}$ is the number operator at site $j$, while $U$ is the strength of the on-site interaction. 
Current experimental feasibilities in arrays of superconducting qubits allow for a high degree of control of these~\cite{roushan17,Martinis2018}. In this platform, one can achieve strong interactions between microwave photons that enables one to perform quantum simulation of condensed matter systems. For example, a recent experiment has shown spectral signatures of localization and ergodicity~\cite{roushan17}. In this work we investigate what happens when a localized system couples to a driven system that is ergodic, to form a ergodic-localized (EL) junction.

Intuition tells us that if the drive is strong enough, it should overcome the effect of disorder and create delocalized states along the whole lattice. In this case, there is a proximity effect: the ergodic system influences the localized one. Further, the disorder within one domain, influences the ergodic behavior of the other domain. To investigate this, we consider a partition of the lattice Eq.~\eqref{eq:1DBosonicGmonArray} into two domains with $M$ sites each, such that $L=2M$. By using this representation, the total system~\eqref{eq:1DBosonicGmonArray} can be decomposed in terms of a localized system $\hat{H}_{\text{Loc}}$ coupled to an ergodic ``bath" $\hat{H}_{\text{Erg}}(t)$, as follows
%%%
\begin{equation}
\label{eq:SystemBath}
\hat{H}(t)=\hat{H}_{\text{Loc}}+\hat{H}_{\text{Erg}}(t)+\hat{H}_{\text{Int}}(t)
\ .
\end{equation}
%%%
Here the interaction between the localized and ergodic domains is given by $\hat{H}_{\text{Int}}(t)=\hbar g(t)(\hat{a}^{\dagger}_{M}\hat{a}_{M+1}+\hat{a}_{M}\hat{a}^{\dagger}_{M+1})$. 
In our case, however, the ``bath" is a domain of the lattice that is externally driven. We do not assume a priori, that the domain is in a thermal state, but consider that it becomes ergodic due to the drive~\cite{Polkovnikov2013,Moessner2014,Rigol2014}.

%%%%%%%%%%%%%%%%%%%%%%%%%%%%%%%%%%%%%%%%%%%%
%\textit{The localized domain: The effect of on-site disorder.\textemdash}
%%%%%%%%%%%%%%%%%%%%%%%%%%%%%%%%%%%%%%%%%%%%
Before we investigate the dynamics of the EL junction, let us discuss the main features of the individual domains. We begin by defining the Hamiltonian of the localized domain as
%%%
\begin{equation}
\label{eq:LocalizedDomain}
\hat{H}_{\text{Loc}}=\hbar\sum^{M}_{j=1}\left[h_{j} \hat{n}_j+\frac{U}{2}\hat{n}_{j}(\hat{n}_{j}-1)\right]+\hbar g_0\sum^{M-1}_{j=1}(\hat{a}^{\dagger}_j\hat{a}_{j+1}+\text{H.c})
\ ,
 \end{equation}
 %%%
 which is a sublattice with sites $j=1,2,\ldots,M$.
For the purposes of this work, we are interested in the spatial dependence $h_{j}=h\cos(2\pi j/M)+\Delta_j$. Here, $\Delta_j\in[-W,W]$ denotes the disorder drawn from a uniform distribution with strength $W$.
We also consider a time-independent coupling $g_0$ between neighboring sites.
 Let us discuss first the physics of the clean system ($W=0$) within the single-particle manifold $(N=1)$, where the interaction term proportional to $U$ does not play a role.
As a consequence of the spatial profile of the on-site energies, the single-particle states are localized between the energy branches $E_{\pm}(j)/\hbar=\frac{h}{2}\cos\left(\frac{2\pi j}{M}\right)\mp2g_0$, a phenomenon known as Bragg localization~\cite{Quintanilla2004, Zoller2017}. $E_{+}(j)$ and $E_{-}(j)$ give us information about the classical (for long wavelength modes) and Bragg (short wavelength modes) turning points~\cite{Quintanilla2004}, respectively.
However, when the disorder is stronger than the coupling $g_0$, the states are localized due to Anderson localization~\cite{anderson58, kramer93,Mirlin2008}. In the case of $N\geq2$ particles, even if the single-particle states are localized, the interaction $U$ may lead to delocalized states~\cite{Altshuler2006, nandkishore15,Vosk2015,khemani17} as it was observed experimentally using two interacting photons in an array of nine superconducting qubits~\cite{roushan17}.
%%%%
%%%%
\begin{figure}
\includegraphics[scale=0.13]{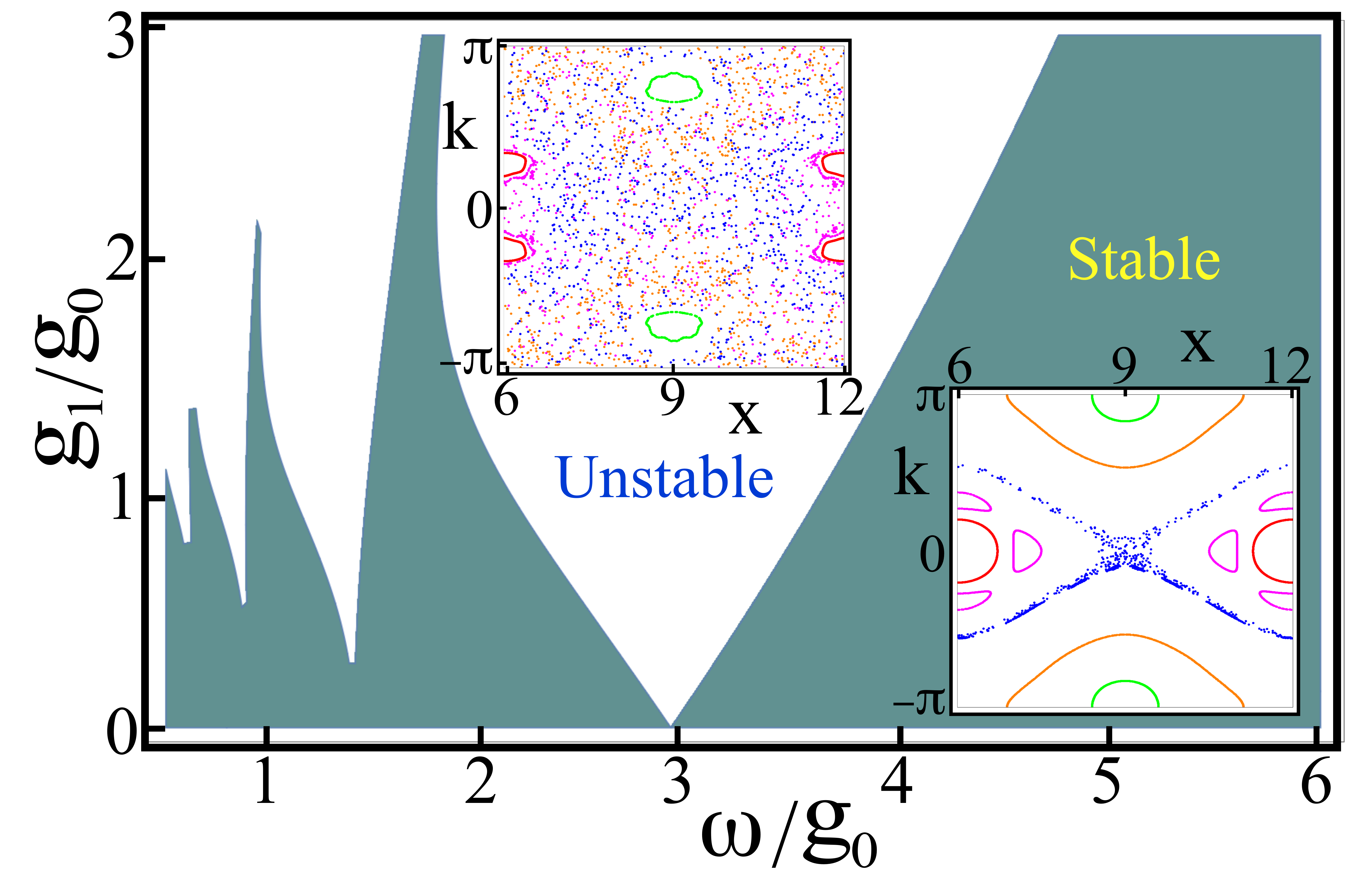}
\vspace{-0.5cm}
\caption{Stability diagram for the classical fixed points $(x,k)=(M,0)$ of the ergodic domain. Parametric resonance occurs at frequencies $m\omega=2\Omega_0$ with integer $m$. Here $\Omega_0$ is the frequency of small oscillations around the fixed point and $\omega$ is the driving frequency. The coloured and white regions depict the stable and unstable zones, respectively. The insets show Poincare maps obtained from Eq.\eqref{eq:SinParClassHam} for parameters within the stable $(\omega,g_1)=(5g_0,0.9g_0)$ and unstable $(\omega,g_1)=(2\Omega_0,0.9g_0)$ zones. We set $h=g_0$, and $M=6$ such that $\Omega_0=1.48g_0$.}
\label{Fig1}
\end{figure}

%%%%
%%%%

%%%%%%%%%%%%%%%%%%%%%%%%%%%%%%%%%%%%%%%%%%%%
%\textit{Ergodic domain: Parametric resonance and Hamiltonian chaos.\textemdash}
%%%%%%%%%%%%%%%%%%%%%%%%%%%%%%%%%%%%%%%%%%%%
Let us now explore in detail the most relevant aspects of the ergodic domain. In particular, we will explain the subtle relation between parametric resonance~\cite{1997kohler,Brandes2010} and the emergence of ergodicity~\cite{Polkovnikov2013}.
The Hamiltonian describing the ergodic domain of the bosonic lattice~\eqref{eq:1DBosonicGmonArray} can be written as
%%%
\begin{equation}
\label{eq:ErgodicDomain}
\hat{H}_{\text{Erg}}(t)=\hbar\sum^{L}_{j=M}\left[\tilde{h}_j  \hat{n}_j+\frac{U}{2}\hat{n}_{j}(\hat{n}_{j}-1)\right]+\hbar g(t)\sum^{L-1}_{j=M+1}(\hat{a}^{\dagger}_j\hat{a}_{j+1}+\text{H.c}),
 \end{equation}
 %%%
with sites $j=M,M+1,\dots,L$ and $\tilde{h}_j=h\cos\left(2\pi j /M\right)$. Within this domain, we ignore the effects of disorder ($W=0$), and consider a time-dependent coupling $g(t)=g_0+g_1\cos(\omega t)$ between the qubits. An external drive can lead to parametric resonance~\cite{1997kohler,Brandes2010}. Within the single-particle manifold and by following the same procedure as in Ref.~\cite{Zoller2017}, one can obtain the  classical Hamiltonian
 \begin{equation}
 \label{eq:SinParClassHam}
 \mathcal{H}(x,k,t)=h\cos\left(\frac{2\pi x}{M}\right) \textcolor{blue}{+}2g(t)\cos k
 \ .
 \end{equation}
Here the kinetic energy is nonlinear, and the particle moves in a cosine potential. From this
we obtain the frequency  $\Omega^2_0=8\pi^2h g_0/M^2$ of the periodic orbits surrounding the classical equilibrium position  
$(x,k)=(M,0)$. 
The phenomenon of parametric resonance appears when the driving frequency $\omega$ is twice the frequency of small oscillations $\Omega_0$~\cite{1997kohler,Brandes2010}.   
Due to our drive, the oscillations around the equilibrium position become unstable for $g_1>0$. When the drive is strong enough $g_1\approx g_0$, most of the regular structures in phase space disappear and the system becomes fully chaotic~\cite{Polkovnikov2013}. Fig.~\ref{Fig1} show the stability of the fixed point $(x,k)=(M,0)$ and the emergence of parametric resonance and chaos. From now on, we drive the system with the resonance condition $\omega=2\Omega_0$.

%%%%%%%%%%%%%%%%%%%%%%%%%%%%%%%%%%%%%%%%%%%%
%\textit{The ergodic-localized (EL) junction: Floquet theory and the stroboscopic dynamics\textemdash}
%%%%%%%%%%%%%%%%%%%%%%%%%%%%%%%%%%%%%%%%%%%%
Let us now explore what happens when the localized and ergodic domains are coupled via the interaction Hamiltonian  $\hat{H}_{\text{Int}}(t)$. One of the most natural questions to ask is to which extent the localized domain is stable when it is coupled to the ergodic one in the case of $N\geq 2$ interacting particles. This resembles a common situation in the theory of open quantum systems: a system is coupled to a thermal bath at a given temperature. In that context, one would expect that if the system-bath coupling is weak, and if the correlation time of the bath is very short, the system thermalizes~\cite{Brandes2008}. Of course, there are some caveats in this argument arising from symmetries preventing thermalization~\cite{Brandes2008}. Symmetries are responsible for level crossings in the spectrum, and the system fails to reach a diagonal ensemble in the long-time limit~\cite{Polkovnikov2013,Moessner2015,Abanin2015,Moessner2016}.

To unveil the dynamics of the EL junction, we invoke Floquet theory for time-periodic Hamiltonians~\cite{Haenggi1998}. This is a natural choice because the Hamiltonian~\eqref{eq:1DBosonicGmonArray} of the total system is periodic, that is, $\hat{H}(t+T)=\hat{H}(t)$, where $T=2\pi/\omega$ is the period of the drive. We will now use the Floquet operator $\hat{\mathcal{F}}=\hat{U}(T)$, which is the evolution operator $\hat{U}(t)$ in one period of the drive~\cite{Haenggi1998,Polkovnikov2015,Anisimovas2015}. The most relevant information can be obtained by solving the eigenvalue problem $\hat{\mathcal{F}}\ket{\Phi_{\mu}}=e^{-\mathrm{i}\varepsilon_{\mu}T/\hbar}\ket{\Phi_{\mu}}$. The eigenvectors $\ket{\Phi_{\mu}}$ are known as the Floquet states and $-\hbar\omega/2\leq\varepsilon_{\mu}\leq\hbar\omega/2$ are the quasienergies.  At discrete times $t_n=nT$, an initial state $\ket{\Psi(0)}$ evolves stroboscopically as $\ket{\Psi(nT)}=\hat{\mathcal{F}}^n\ket{\Psi(0)}$. This motivates the use of the effective Hamiltonian $\hat{H}_{\text{Eff}}$: a generator of the stroboscopic dynamics $\hat{\mathcal{F}}=e^{-\mathrm{i}\hat{H}_{\text{Eff}}T/\hbar}$. It gives us important information of the effective interactions that appear due to the drive and it can be interpreted in terms of quantum simulation~\cite{Polkovnikov2015,Brandes2016,Eckardt2017}. Nonetheless, it is a difficult task to obtain $\hat{H}_{\text{Eff}}$ analytically.  As a matter of fact, there are high-frequency expansions~\cite{Polkovnikov2015,Anisimovas2015} that allow one to obtain analytical expressions up to a finite order in $\omega^{-1}$. In this work we are interested in the low-frequency regime, where all the high frequency expansions are known to diverge~\cite{Saito2016}. Our approach to solve this problem is to numerically calculate the effective Hamiltonian by taking the logarithm of the Floquet operator $-\mathrm{i}\hat{H}_{\text{Eff}}T/\hbar=\log_e\hat{\mathcal{F}}$. To obtain the matrix representation of the effective Hamiltonian, we consider a basis  $\ket{l}$ of all the possible $D_N=(L+N-1)!/N!((L-1)!$ configurations of $N$ particles distributed in $L$ lattice sites, where $l=1,\ldots,D_N$ (see Appendix~\ref{Sec.I}). In the case $N=2$ and $L=12$ that we consider in this manuscript, there are $D_2=78$ configurations $\ket{l}=\ket{0,\ldots,1_i,\ldots,1_j}$. All the information we are interested in is contained in the matrix elements $(H_{\text{Eff}})_{l,\tilde{l}}$ of the effective Hamiltonian, which can be used to visualize the total system as a graph with $D_2$ nodes using TULIP5~\cite{auber2017tulip}, and further allows us to determine localization properties of the individual domains. In addition, one can represent the matrix as a graph, which allows us to unveil the formation of clusters and communities~\cite{Goodger2012}. In so doing, we construct the adjacency matrix $\mathcal{A}$ by following the rules $\mathcal{A}_{l,l}=0$ and $\mathcal{A}_{l,\tilde{l}}=1$ if $|(H_{\text{Eff}})_{l,\tilde{l}}|>C$, where $C=10^{-2}g_0$ is a cutoff that we introduce to have a better visualization. 
%%%%
%%%%
\begin{figure}
\includegraphics[scale=0.36]{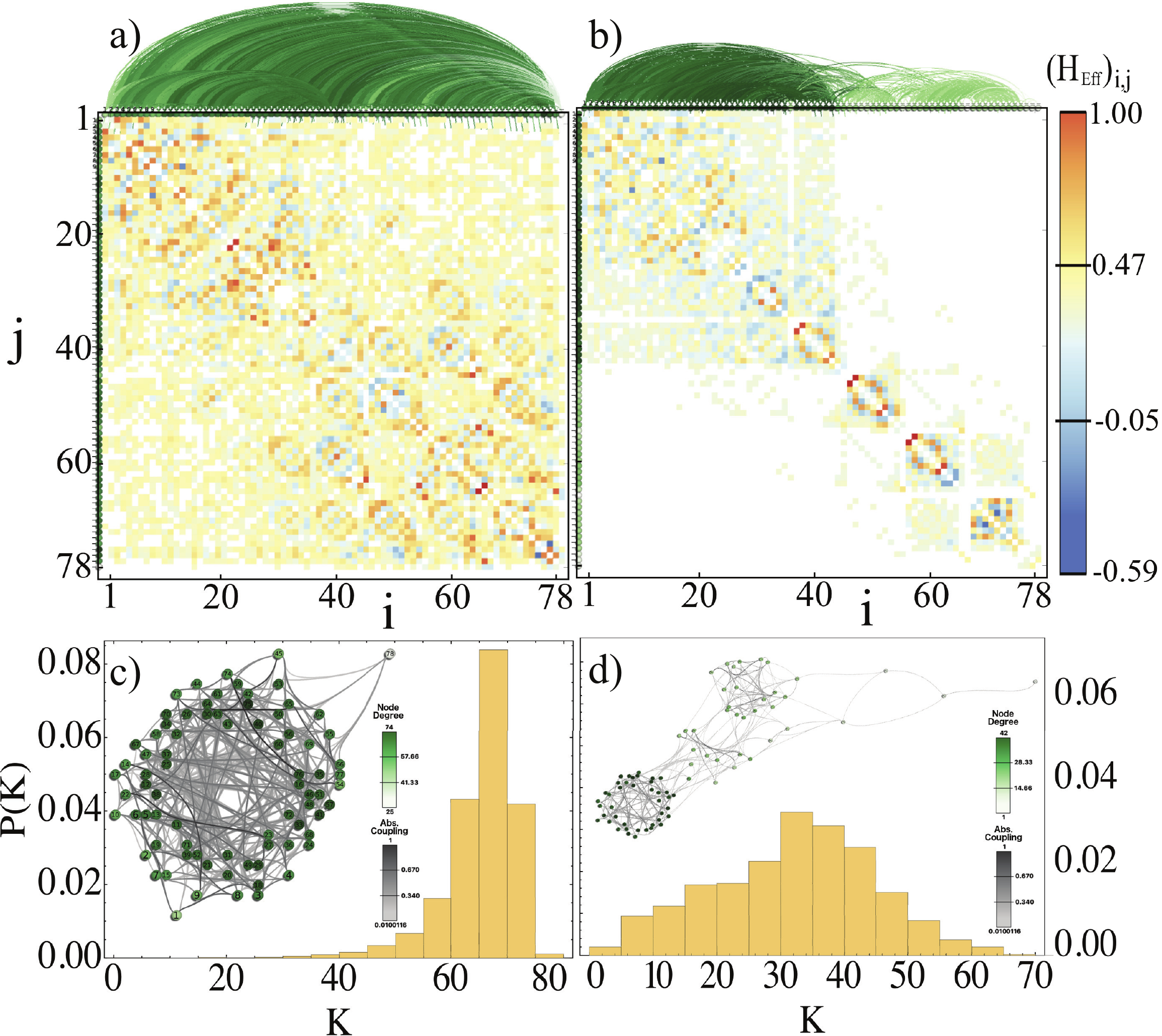}
\vspace{-0.5cm}
\caption{Visual representation of the effective Hamiltonian for two interacting particles as a graph. (a) and (b) depict the effective Hamiltonian for $W=g_0$ and $W=10g_0$, respectively. On top of the matrices, we show the graph connectivity between the different sites. Next, c) and d) show the degree distribution $P(K)$ obtained from 100 realizations of disorder for $W=g_0$ and $W=10g_0$, respectively. The insets show the formation of clusters in the graphs associated to the matrices shown in a) and b).
The total system consists of two interacting particles in a one-dimensional lattice with $L=12$ sites, where each domain has $M=6$ sites.
We set $h=g_0$, $U=3.5g_0$ and  a drive with frequency $\omega=2\Omega_0=2.96g_0$ and amplitude $g_1=0.9g_0$.}
\label{Fig2}
\end{figure}
%%%%
%%%
 
%%%%
%%%%
\begin{figure}
\includegraphics[scale=0.112]{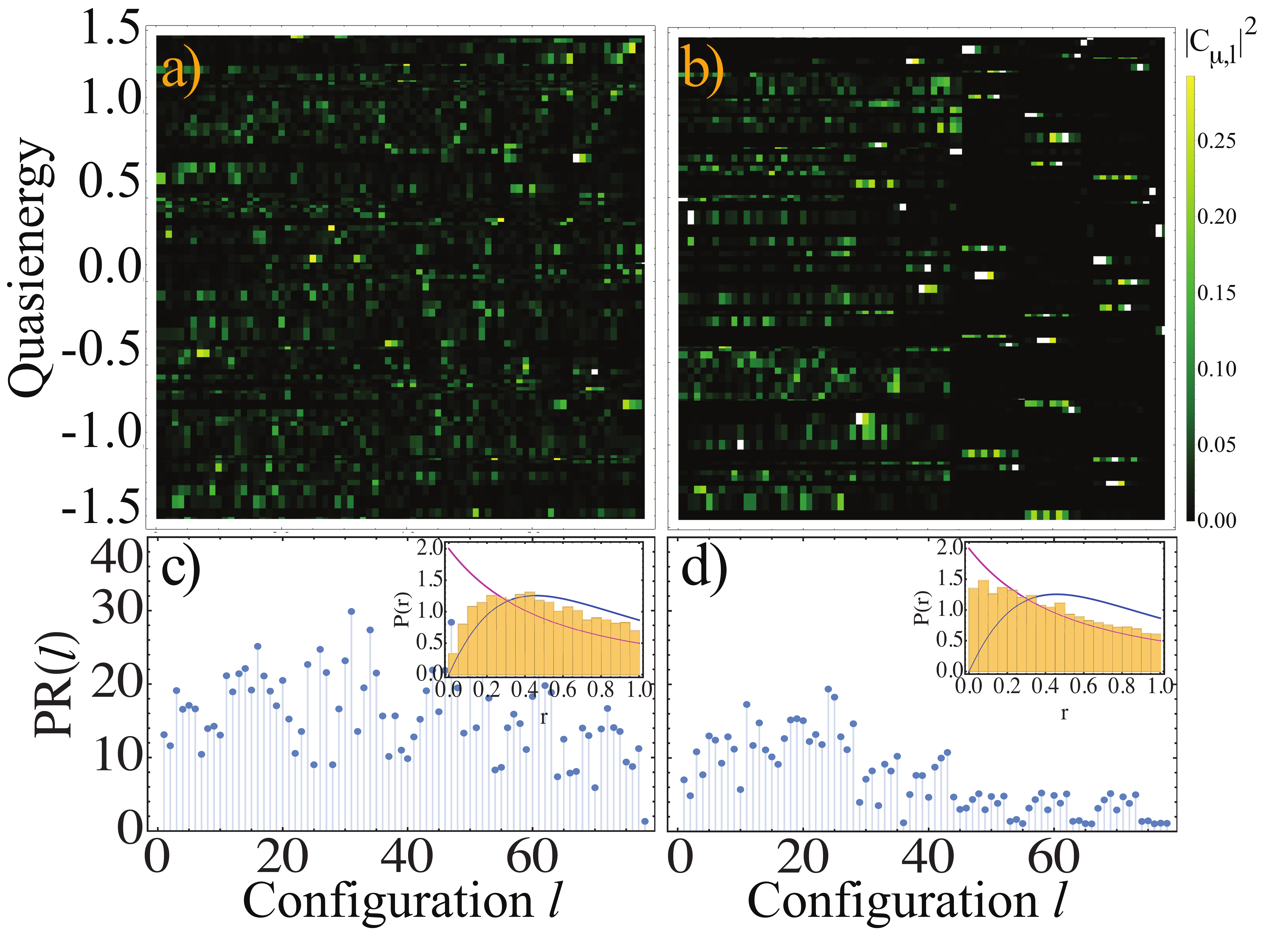}
\vspace{-0.65cm}
\caption{
Spatial localization of the Floquet states and Participation ratio for $N=2$ interacting bosonic particles in $L=12$ sites. For a single realization of disorder, Figs. (a) and (b) depict the probability amplitude $|c_{\mu,l}|^2$ for $W=g_0$ and $W=10g_0$, respectively. The latter is shown as a function of the configuration $l$ and the quasienergy value $\varepsilon_{\mu}$. Figs. (c) and (d) show the participation ratio $PR(l)$ for $W=g_0$ and $W=10g_0$, respectively. The insets in  Figs. (c) and (d) show the level statistics of the quasienergies for 100 realizations of disorder, which are very close to GOE (blue curve) and Poisson (magenta curve), respectively. For a disorder strength $W=g_0$, one can observe the proximity effect: the localized domain becomes unstable due to the coupling to the ergodic domain.
 We set $h=g_0$, $U=3.5g_0$, and a drive with a frequency $\omega=2\Omega_0=2.96g_0$ and amplitude $g_1=0.9g_0$.}
\label{Fig3}
\end{figure}
%%%%
%%%%
Figure~\ref{Fig2} shows the matrix representation of the effective Hamiltonian for a single realization of disorder. There, one can recognize two domains with different behavior. When the disorder is moderate, $W=g_0$, there is a proximity effect and a region close to the interface becomes ergodic. When the disorder is strong, $W=10g_0$, the localized domain is stable and fails to become ergodic. Note that in general, as depicted in Figs.~\ref{Fig2}~(a) and~(b), the connectivity of the graph is high in the ergodic domain and low in the localized one. In figure~(c) there is a formation of two clusters in the graph. Due to the proximity effect, there is certain connectivity between these two clusters. In contrast, although figure~(d) shows two clusters, the connectivity is low between them. One way to quantify the connectivity is to calculate the degree $K$ of a node, which is the number of links to other nodes in the graph~\cite{Mendes2008}.  To obtain the probability distribution $P(K)$, which gives us the probability of a node to have a degree $K$~\cite{Mendes2008}, we consider $100$ realizations of disorder.  In Fig.~\ref{Fig2}~(c) one can see that $P(K)$ has small variance and high average degree in the case of moderate disorder and Fig.~\ref{Fig2}~(d) shows that the variance is bigger with low average degree for strong disorder. The connectivity of the ergodic domain resembles the emergence of a giant component and percolation in random networks~\cite{Mendes2008}.

%%%%%%%%%%%%%%%%%%%%%%%%%%%%%%%%%%%%%%%%%%%%
%\textit{Localization properties of the states.\textemdash}
%%%%%%%%%%%%%%%%%%%%%%%%%%%%%%%%%%%%%%%%%%%%
Now let us study localization properties of the Floquet states in the case of two particles in $L=12$ sites by using the participation ratio~\cite{kramer93,roushan17}. 
This quantity measures how localized is a state in a given basis.
Any Floquet state can be decomposed as a quantum superposition of configurations $\ket{l}$, as follows $\ket{\Phi_{\mu}}=\sum^{D_N}_{l=1} c_{\mu,l}\ket{l}$. From this decomposition one can see that the number of coefficients $c_{\mu,l}$ determine how extended is $\ket{\Phi_{\mu}}$ in the basis $\ket{l}$. Conversely, one can also quantify how extended is a configuration in the energy eigenbasis
%%%
\begin{equation}
\label{eq:ParticipationRatioE}
        PR(l)={1}/{\sum^{D_N}_{l=1} |c_{\mu,l}|^4}
        \ .
\end{equation}
%%%
The participation ratio measures how many states ``participate" in a quantum superposition.
One might be tempted to interpret the participation ratio as a localization length, but this is not correct. In fact, the participation ratio is commonly interpreted as the ``Radius" of the wave function~\cite{kramer93}. In contrast, the localization length has to do with the rate of spatial exponential decay of the state along the lattice and plays an important role in the theory of Anderson and many-body localization~\cite{kramer93}.  
For single realization of disorder, Figs.~\ref{Fig3}(a) and (b) show the probability amplitudes $|c_{\mu,l}|^2$ as a function of the configuration $l$ and the quasienergies $\varepsilon_{\mu}$ for moderate ($W=g_0$) and strong disorder ($W=10g_0$), respectively. This figure nicely reflects the geometrical representation shown in figure~\ref{Fig2}. In addition to this, Figs.~\ref{Fig3}(c) and (d) show the participation ratio for a single realization of disorder. The high connectivity between nodes in the ergodic domain is related to localization properties of the Floquet states. In the case of strong disorder, the localized domain remains stable in despite of being coupled to the ergodic one. For weaker disorder, the proximity effect is stronger and the states becomes delocalized. When the disorder of order $W<g_0$, most of the states become delocalized. Furthermore, the effective Hamiltonian is highly non-local and the clusters of the associated graph disappear, i.e., the graph becomes almost fully connected. In the Appendix~\ref{Sec.III}, we show results for $N=1$ and $N=3$ that show a similar behavior to the one described above. To have an idea of how ergodic or localized is the system, we resort on statistical properties of the quasienergies, also referred to as level statistics~\cite{Rigol2014,roushan17}. To obtain this, we consider an ordered sequence of quasienergies $\varepsilon_1<\varepsilon_2,\ldots,\varepsilon_{D_N-1}<\varepsilon_{D_N}$. Based, on this, we define the nearest-neighbor spacings $s_\mu=\varepsilon_{\mu+1}-\varepsilon_{\mu}$ and the ratio $r_{\mu}=\text{min}(s_\mu,s_{\mu-1})/\text{max}(s_\mu,s_{\mu-1})$. The insets of Figs. (c) and (d) depict the statistics of $\{r_{\mu}\}$ obtained from $100$ realization of disorder (see Appendix~\ref{SecIIF}). From this one can see that when there is a proximity effect, the statistics is close to the Gaussian Ortogonal ensemble (GOE). Correspondingly, in the case of a stable localized domain, the statistics is very close to a Poissonian distribution~\cite{Rigol2014,roushan17}.
%%%%%%%%%%%%%%%%%%%%%%%%%%%%%%%%%%%%%%%%%%%%
%\textit{Conclusions.\textemdash}
%%%%%%%%%%%%%%%%%%%%%%%%%%%%%%%%%%%%%%%%%%%%

To summarize, in this letter we explored the situation where one couples two domains of a system of $N$ interacting bosons in a one-dimensional lattice with $L$ sites: one of them is ergodic due to the external drive and the other one is fully localized. The total system constitutes an ergodic-localized (EL) junction. By using Floquet theory, we show that the localized domain remains stable for strong disorder. When the disorder is decreased, there is a proximity effect and the localized domain becomes ergodic. We provide a geometrical interpretation of this phenomenon by representing the effective Hamiltonian as a graph. The connectivity of the associated graph reveals if certain domains of the system are localized or ergodic. This behavior can be quantified by using the degree distribution $P(K)$ that has a small variance with a high mean value when the localized domain is unstable.
Possible implementations of our results could be achieved using existing quantum technologies such as cold atoms~\cite{Bloch2015,Gross2016}, trapped ions~\cite{Monroe2016} and superconducting qubits arrays (Appendix~\ref{Sec.V})~\cite{roushan17,Martinis2018}, where one has control of the on-site energies $h_{j}$, that can be tuned to define different spatial profiles and the coupling strengths $g_j(t)$ can be modulated in certain domains of the array. 
 We anticipate that our work will open a new avenue of research and inspire the use of graph theory to unveil the dynamics of periodically-driven quantum systems. A possible application of our approach is to perform stroboscopic quantum simulation~\cite{Polkovnikov2015, Brandes2016,Eckardt2017} of complex network topologies\cite{Bandyopadhyay2007} by driving a system of qubits with a simple topology (see Appendix~\ref{Sec.IV}). Furthermore, our methodology should be generalizable to investigate periodic random circuits~\cite{Cirac2018}, and many-body localized states such as time crystals~\cite{else16,yao17,Zakrzewski2018}.

%%%%%%%%%%%%%%
%\section{Acknowledgement}
%%%%%%%%%%%%%%
We thank K. Azuma, M. Hanks, T. Haug, F. Katsuya, S. Restrepo, P. Roushan  and J. Tangpanitanon for fruitful discussions. This work was supported in part by the MEXT KAKENKHI Grant number No 15H05870 and through the support of a John Templeton Foundation grant (JTF No 60478). The opinions expressed in this publication are those of the authors and do not necessarily reflect the views of the John Templeton Foundation.

\appendix

%%%%%%%%%%%%%%%%%%%%%%%%%%%%%%%%%%%%%%%%%%%%%%%%%%%%%%
%%%%%%%%%%%%%%%%%%%%%%%%%%%%%%%%%%%%%%%%%%%%%%%%%%%%%%
\section{Basis of the Hilbert space\label{Sec.I}}
%%%%%%%%%%%%%%%%%%%%%%%%%%%%%%%%%%%%%%%%%%%%%%%%%%%%%
In our manuscript, we focus on a system of $N$ interacting bosons in a one-dimensional lattice with $L$ sites given by Hamiltonian Eq.$[1]$. Due to the statistics of the particles, the basis $\ket{l}$ of the Hilbert space consists of $D_N=(L+N-1)!/N!((L-1)!$ configurations of $N$ particles distributed in $L$ lattice sites, where $l=1,\ldots,D_N$. Mathematically, these configurations correspond to all the possible compositions of $N$ particles into $L$ parts. In the cases of $N=1$, $N=2$ and $N=3$ particles in $L=12$ sites, the basis has dimensions $D_1=12$, $D_2=78$ and $D_3=364$, respectively. 
%%%%
\begin{figure}[h]
\includegraphics[scale=0.63]{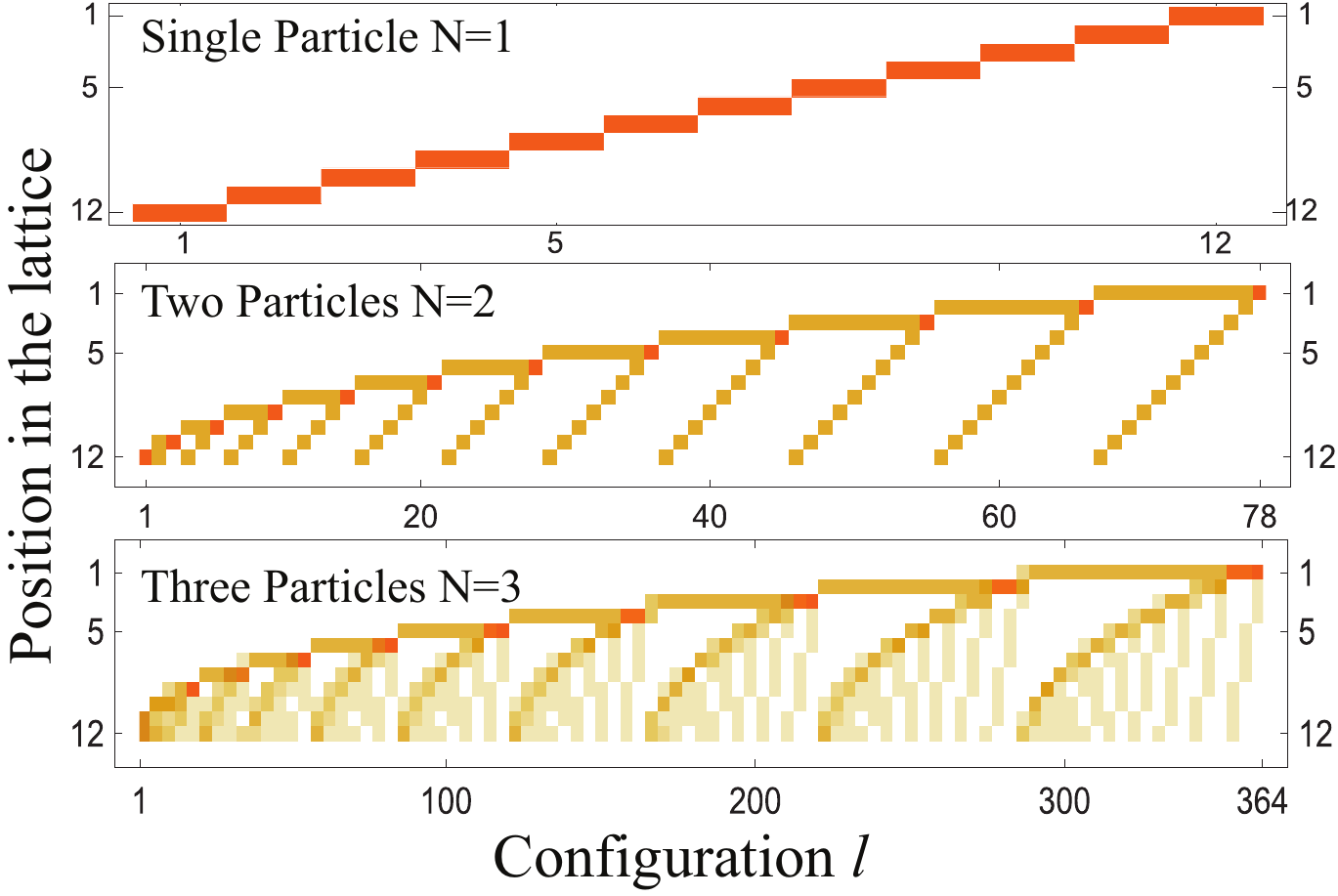}
\vspace{-0.1cm}
\caption{Visual representation of basis of the Hilbert space for $N$ interacting particles in $L$ sites. To construct the basis, we consider compositions of $N=1,2,3$ particles in $L=12$ lattice sites. 
}
\label{Fig0S}
\end{figure}
%%%%
Fig.~\ref{Fig0S} depicts a visual representation of these basis. For example, in the case of $N=3$ particles, $\ket{1}=\ket{0,0,\dots,0,3}$, $\ket{2}=\ket{0,0,\dots,1,2}$, $\ket{3}=\ket{0,0,\dots,2,1}$ and $\ket{4}=\ket{0,0,\dots,3,0}$ are the first four configurations. In this example, the last configuration is $\ket{364}=\ket{3,0,\dots,0,0}$. Each one of basis elements constitutes a node of the graphs discussed here and in our manuscript.

%%%%%%%%%%%%%%%%%%%%%%%%%%%%%%%%%%%%%%%%%%%%%%%%%%%%%%
\section{Floquet theorem applied to the Schr\"odinger equation and statistical behavior of the quasienergies\label{SecIIF}}
%%%%%%%%%%%%%%%%%%%%%%%%%%%%%%%%%%%%%%%%%%%%%%%%%%%%%
In the case of time-periodic Hamiltonian $\hat{H}(t)=\hat{H}(t+T)$, there is not stationary states and one faces the
solution of the time-dependent Schr\"odinger equation $\mathrm{i}\hbar\partial_t\ket{\Psi(t)}=\hat{H}(t)\ket{\Psi(t)}$. In this case, however, we can use Floquet theory~\cite{Haenggi1998}, which is based on the ansatz
%%%%
\begin{equation}
      \label{eq:BasisExp}
            \ket{\Psi(t)}=\sum_{\lambda}c_{\lambda}e^{-\frac{\mathrm{i}}{\hbar}\varepsilon_{\lambda}t}\ket{\Phi_{\lambda}(t)}
      \ ,
\end{equation}
%%%%
where $\ket{\Phi_{\lambda}(t)}= \ket{\Phi_{\lambda}(t+T)}$ are the Floquet modes, and $\varepsilon_{\lambda}$ the quasienergies. In the case of Hamiltonian Eq.$[1]$, once we have chosen a basis $\ket{l}$ for the Hilbert space of $N$ interacting particles in $L$ sites, the Schr\"odinger equation can be written as a linear system of $D_N$ coupled ordinary differential equations
$\mathrm{i}\hbar\frac{d \boldsymbol{\Psi} (t)}{d t}=\boldsymbol{{H}}(t)\cdot\boldsymbol{\Psi} (t)$. Here, $\boldsymbol{\Psi} (t)=[\psi_1(t),\ldots,\psi_{D_N}(t)]$ is the vector representation of the quantum state $\ket{\Psi(t)}$ with $\psi_l(t)=\langle l\ket{\Psi(t)}$. In addition, the $D_N\times D_N$  time-periodic matrix $\boldsymbol{H}(t+T)=\boldsymbol{H}(t)$ is the matrix representation of the Hamiltonian $\hat{H}(t)$ in the basis $\ket{l}$, i.e., $H_{l,\tilde{l}}(t)=\bra{l}\hat{H}(t)\ket{\tilde{l}}$. 

%%%%%%%%%%%%%%%%%%%%%%%%%%%%%%%%%%%%%%%%%%%%%%%%%%%%%%
\subsection{Floquet theorem and the Schr\"odinger equation}
%%%%%%%%%%%%%%%%%%%%%%%%%%%%%%%%%%%%%%%%%%%%%%%%%%%%%
The solutions of the system of differential equations discussed above are not periodic, but the Floquet theorem enables us to obtain important information about them~\cite{1883floquet, 1975yakubovich}. Let us define a square matrix $\boldsymbol{U}(t)$ whose columns are $D_N$ solutions $\{\boldsymbol{\Psi}_1 (t),\ldots, \boldsymbol{\Psi}_{D_N} (t)\}$ of the system of differential equations. $\boldsymbol{U}(t)$ is the matrix representation of the evolution operator $\hat{U}(t)$, which is also known as the fundamental matrix in the theory of differential equations~\cite{1975yakubovich}.  Due to the periodicity of the Hamiltonian, it can be shown that $\boldsymbol{U}(t+T)=\boldsymbol{U}(t)\cdot\boldsymbol{F}$, where $\boldsymbol{F}=\boldsymbol{U}(T)=e^{-\frac{\mathrm{i}}{\hbar}\boldsymbol{H}_{\text{Eff}}T}$ is the matrix representation of the Floquet operator and $\boldsymbol{H}_{\text{Eff}}$ is the effective Hamiltonian generating the stroboscopic dynamics. The eigenvalues $e^{-\frac{\mathrm{i}}{\hbar}\varepsilon_\mu T}$ of $\boldsymbol{F}=\boldsymbol{U}(T)$ are characteristic multipliers. The arguments $\varepsilon_\mu$ of these eigenvalues are the Floquet exponents or quasienergies and due to its form, they not uniquely defined~\cite{Haenggi1998}. For example, if $\varepsilon_\mu $ is a Floquet exponent, then $\varepsilon_\mu +\hbar\omega n$ with integer $n$ and $\omega=2\pi/T$, is also a Floquet exponent. 
%%%%
%%%%
\begin{figure*}
\includegraphics[scale=0.11]{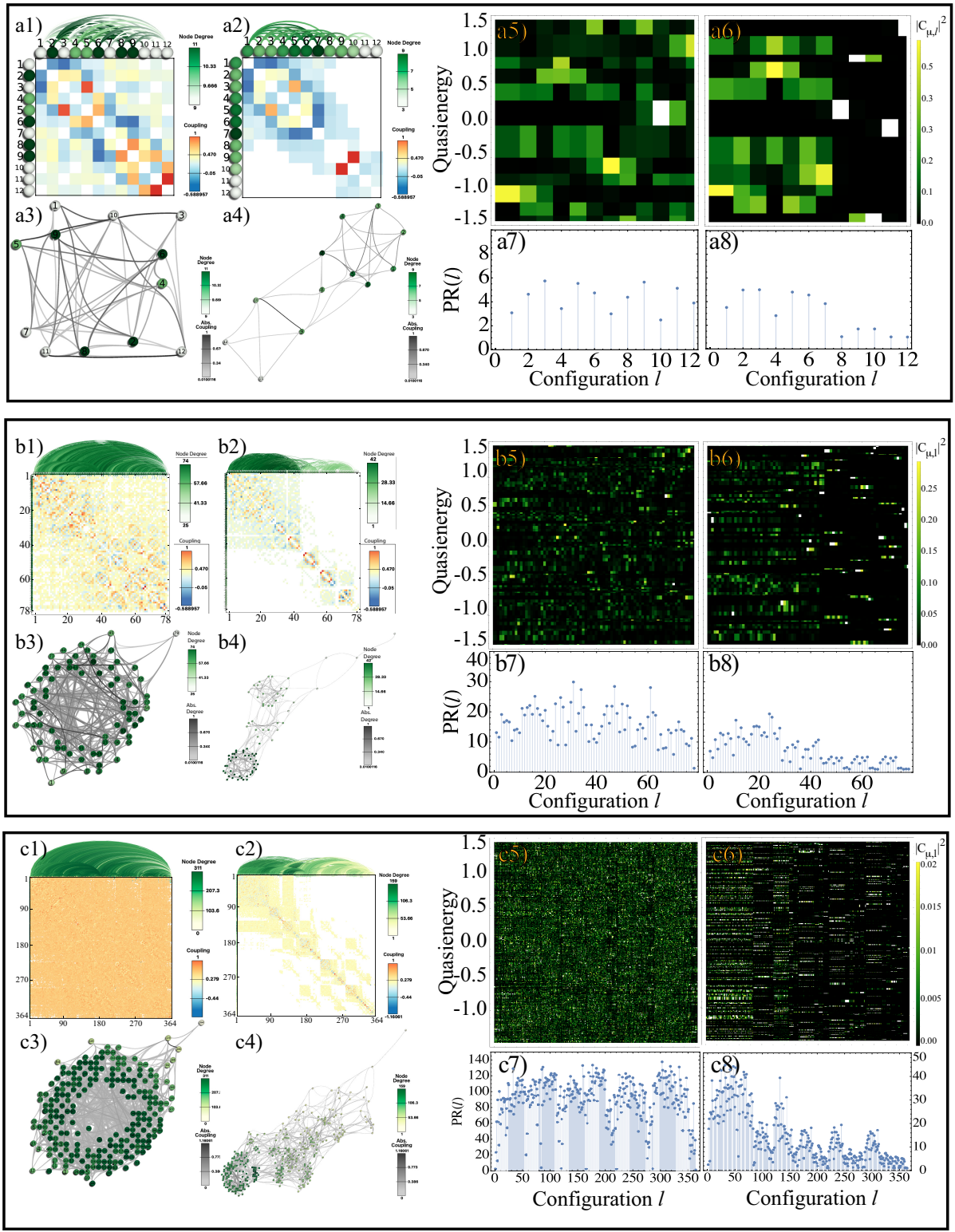}
\vspace{-0.1cm}
\caption{Visual representation of the effective Hamiltonian for a) $N=1$, b) $N=2$ and c) $N=3$ interacting particles as a graph and spatial localization of the Floquet states  for a single realization of disorder. The panels [a1), b1), c1)] and [a2), b2), c2)] depict the effective Hamiltonian for  $W=g_0$ and $W=10g_0$, respectively. On top of the matrices, we show the graph connectivity between the different sites. The panels [a3), b3), c3)] and [a4), b4), c4)] show the graphs associated to the matrices shown in [a1), b1), c1)] and [a2), b2), c2)] .  Figs. [a5), b5), c5)] and [a6), b6), c6)] depict the probability amplitude $|c_{\mu,l}|^2$ for $W=g_0$ and $W=10g_0$, respectively. The latter is shown as a function of the configuration $l$ and the quasienergy value $\varepsilon_{\mu}$. Figs. [a7), b7), c7)] and [a8), b8), c8)]  show the participation ratio corresponding to Figs. [a5), b5), c5)] and [a6), b6), c6)] .
The total system consists of $N=1,2,3$ particles in a one-dimensional lattice with $L=12$ sites, where each domain has $M=6$ sites.
We set $h=g_0$, $U=3.5 g_0$ and  a drive with frequency $\omega=2\Omega_0=2.96g_0$ and amplitude $g_1=0.9g_0$.}
\label{Fig1MS}
\end{figure*}
%%%%
%%%%
%%%%%%%%%%%%%%%%%%%%%%%%%%%%%%%%%%%%%%%%%%%%%%%%%%%%%
\subsection{Level statistics for the quasienergies}
%%%%%%%%%%%%%%%%%%%%%%%%%%%%%%%%%%%%%%%%%%%%%%%%%%%%%
As we have discussed previously, the quasienergies are not unique. For this reason, we have to restrict them to the first Brillouin zone $-\hbar\omega/2\leq\varepsilon_{\mu}\leq\hbar\omega/2$~\cite{Haenggi1998}. After this, we can construct an ordered sequence  $\varepsilon_1<\varepsilon_2,\ldots,\varepsilon_{D_N-1}<\varepsilon_{D_N}$. To investigate the statistical behavior of the quasienergies, we define the nearest-neighbor spacings $s_\mu=\varepsilon_{\mu+1}-\varepsilon_{\mu}$ and the ratio $r_{\mu}=\text{min}(s_\mu,s_{\mu-1})/\text{max}(s_\mu,s_{\mu-1})$, as we discussed in the main text. By considering several realizations of disorder, we can obtain histograms for the data $\{r_{\mu}\}$.
 The probability distributions of $r_{\mu}$ show different behavior depending on the interplay between the interactions and disorder, as it has been observed in a recent experiment in the absence of drive~\cite{roushan17}. When the system is ergodic, the statistical behavior is given by the Gaussian Ortogonal ensemble (GOE). To be more precise, as we are working with quasienergies, the statistical ensemble is the circular ortogonal ensemble (CE), but they are related and in some limits, very close to each other~\cite{Rigol2014}. On the other hand, if the system is localized, the statistics is Poissonian. The corresponding statistical distributions read

\begin{equation}
P_{\text{GOE}}(r)=\frac{27}{4} \frac{r+r^2}{(1+r+r^2)^{5/2}},\,\ P_{\text{Poisson}}(r)=\frac{2}{(1+r)^2}.
\end{equation}
In the main text, we depict the results for the level statistics using $100$ realizations of disorder in the insets of Figs. 3 (c) and (d).

%%%%%%%%%%%%%%%%%%%%%%%%%%%%%%%%%%%%%%%%%%%%%%%%%%%%%%
\section{Representation of the effective Hamiltonian as a graph for $N=1$ and $N=3$ and localization properties of the Floquet states\label{Sec.III}}
%%%%%%%%%%%%%%%%%%%%%%%%%%%%%%%%%%%%%%%%%%%%%%%%%%%%%

The purpose of this section is to show additional results for $N=1,2,3$ particles in a one-dimensional lattice with $L=12$ sites. A useful quantity to characterize an undirected graph is its density, which is defined as the number of edges $|E|$ in comparison to a clique
%%%
\begin{equation}
          \label{eq:Density}
                  D=\frac{2|E|}{|V|(|V|-1)}
          \ ,
\end{equation}
%%%
where $|V|$ is the number of vertices of the graph. A high density is related to more connection in the graph and when the density is one, the graph is a clique. In our manuscript, the number of vertices $|V|=D_N$ is given by the dimension $D_N$ of the $N$-particle manifold.

In the case of a single particle $N=1$, the interactions do not play a role and the configurations are very simple. For example,  $\ket{1}=\ket{0,0,\dots,0,0,1}$,  $\ket{2}=\ket{0,0,\dots,0,1,0}$ and  $\ket{3}=\ket{0,0,\dots,1,0,0}$ are the first three configurations. Figs.~\ref{Fig1MS}(a1) and (a2) show the matrix representation of the effective Hamiltonian and  Figs.~\ref{Fig1MS} (a3), (a4) the associated graphs. The density for the ergodic case is $D=0.89$ and for the localized one is $D=0.56$. This means that even in the ergodic case, the graph is not a clique, but it has clusters with high connectivity. We can see that the connectivity of the graph reflects localization properties of the quantum states as it is shown in Figs.~\ref{Fig1MS}(a5), (a6), (a7) and (a8).  For $N=2$ interacting particles, Figs.~\ref{Fig1MS} b1) and (b2)  depict the effective Hamiltonian and Figs.~\ref{Fig1MS} (b3), (b4) the corresponding graphs. In this case, the graph densities are $D=0.81$ and $D=0.36$ for $W=g_0$ and $W=10g_0$, respectively. Localization properties of the states are shown in Figs.~\ref{Fig1MS}(b5), (b6), (b7) and (b8).
Now we discuss the case $N=3$. Figs.~\ref{Fig1MS}(c1)  and (c2) show the matrix representation of the effective Hamiltonian and (c3), (c4) the corresponding graphs. The densities are $D=0.71$ and $D=0.21$ for weak disorder $W=g_0$ and strong disorder $W=10g_0$, respectively. Interestingly, the density for the localized regime is lower than for $N=1$ and $N=2$. A possible explanation of this is that in the presence of more particles, the interactions tend to localize the states. For completeness, we depict the localization properties of the states in Figs.~\ref{Fig1MS}(c5), (c6), (c7) and (c8).

%%%%%%%%%%%%%%%%%%%%%%%%%%%%%%%%%%%%%%%%%%%%%%%%%%%%%%
\section{Hardcore bosons: the spin representation\label{Sec.IV}}
%%%%%%%%%%%%%%%%%%%%%%%%%%%%%%%%%%%%%%%%%%%%%%%%%%%%%
In this section, we discuss a very interesting limit $U\gg h_j, g_j$ of Hamiltonian Eq. [1]
 in the main text, which is referred to as hardcore bosons regime~\cite{Noh17}.  In this regime one can truncate the local bosonic Hilbert space at a given site $i$ up to two states $\{\ket{0}_i,\ket{1}_i\}$, which allows us to write the low-energy effective Hamiltonian
\begin{align}
\label{eq:1DGmonArraySpins}
\hat{H}(t)&=\frac{\hbar}{2}\sum^{L}_{l=1}h_{l} \hat{Z}_{l}+\frac{\hbar}{2}\sum^{L-1}_{l=1}g_l(t)(\hat{X}_{l}\hat{X}_{l+1}+\hat{Y}_{l}\hat{Y}_{l+1})
 \ .
 \end{align}
Here $\hat{X}_l,\hat{Y}_l,\hat{Z}_l$ are the usual Pauli matrices with $h_l$ and $g_l(t)$ being the strengths of the transverse field and the spin-spin interaction, respectively. In this case, one can use the Jordan-Wigner transformation~\cite{nagaosa1999}

\begin{equation}
 \hat{X}_l = f^{\dagger}_l e^{\mathrm{i}\hat{\Phi}_l}+f_l e^{-\mathrm{i}\hat{\Phi}_l},\,\ \hat{Y}_l = -\mathrm{i}f^{\dagger}_l e^{\mathrm{i}\hat{\Phi}_l}+\mathrm{i}f_l e^{-\mathrm{i}\hat{\Phi}_l}, \,\   \hat{Z}_l =2f^{\dagger}_l f_l-1 \ ,
\end{equation}
with $\hat{\Phi}_l=\sum_{j < l}f^{\dagger}_j f_j$ to map the model to a Hamiltonian of spin-less fermions

%%%
\begin{align}
         \label{eq:Hamiltonian}
                 \hat{H}&=\frac{\hbar}{2}\sum^{N}_{l=1}h_l(2f^{\dagger}_l f_l-1)+\hbar\sum^{N-1}_{l=1}g_l(t)(f^{\dagger}_{l}f_{l+1}+f^{\dagger}_{l+1}f_{l})      \ .
          \end{align}
%%%
Under the effect of periodic driving, we can calculate the effective Hamiltonian in the fermionic representation. After that can we apply the inverse Jordan-Wigner transformation to be able to write it in terms of Pauli matrices, as follows
%%%%
%%%
\begin{equation}
         \label{eq:EffFermionicHamiltonianAnderson}
         \hat{H}_{\text{Eff}}=\hbar\sum^{L}_{l=1}\mathcal{M}_{l,l}\hat{Z}_{l}+\hbar\sum^{L}_{l<\tilde{l}}\mathcal{M}_{l,\tilde{l}}\left(\hat{X}_{l}\hat{O}_{l,\tilde{l}}\hat{X}_{\tilde{l}}+\hat{Y}_{l}\hat{O}_{l,\tilde{l}}\hat{Y}_{\tilde{l}}\right)
                  \ .
          \end{equation}
%%%
The operators
$\hat{O}_{l,\tilde{l}}=\hat{Z}_{l+1}\cdots\hat{Z}_{\tilde{l}-1}$ give rise to
highly non-local terms weighted by the matrix elements $\mathcal{M}_{l,\tilde{l}}=(H_{\text{Eff}})_{l,\tilde{l}}$ that appear due to the Jordan-Wigner strings~\cite{nagaosa1999}.
In the case of a lattice with $L=12$ sites, Figs.~\ref{Fig1MS} (a1) and (a2) show typical matrix elements $\mathcal{M}_{l,\tilde{l}}=(H_{\text{Eff}})_{l,\tilde{l}}$ of effective Hamiltonian for $W=g_0$ and $W=10g_0$, respectively.

%=====================================================================
\section{Experimental protocol to study localization of Floquet states: Spectroscopy method for driven systems\label{Sec.V}}
%=====================================================================@
In diverse communities, one is interested in methods to resolve both the energy spectrum of a system as well as properties of its eigenstates. In a recent experiment using superconducting qubits~\cite{roushan17}, a spectroscopy method has been developed to resolve the energy spectrum of an undriven manybody system. The latter was applied to study localization properties of two interacting photons in nine superconducting qubits. In this section, we extend that method to be able to resolve the localization properties of Floquet states, which is essential to investigate localization phenomena in driven quantum systems such as the Hamiltonian $\hat{H}(t)$ of Eq.[1] that we study in our manuscript. As we discussed in Sec.~\ref{SecIIF}, when the system is periodically driven $\hat{H}(t)=\hat{H}(t+T)$, the quasienergies and the Floquet modes play the role of energies and eigenstates, respectively.

Experimentally, in order to have access to the quasienergy spectrum, we need the expectation values of certain observables at stroboscopic times. The idea behind the method is very similar to the time-independent case presented in Ref.~\cite{roushan17}: we consider an initial state $ \ket{\Psi^{(N)}(0)}=\sum_{\lambda}c_{\lambda}\ket{\Phi^{(N )}_{\lambda}(0)}$ that is a linear superposition of Floquet states $\ket{\Phi^{(N)}_{\lambda}(0)}$ within the $N$-particle manifold. The next step is to define the expectation value $A^{l}_n=\ew{\hat{A}^{l}(nT)}=\bra{\Psi^{(N)}(nT)}\hat{A}^{l}\ket{\Psi^{(N)}(nT)}$ of a given observable $\hat{A}^{l}$ in the state $ \ket{\Psi^{(N)}(nT)}$ at stroboscopic times $t_n=nT$. The observable $\hat{A}^{l}$ is chosen depending on the configuration $\ket{l}$ of $N$ particles in $L$ sites.
For example, to investigate spectroscopic properties of the single-particle manifold $(N=1)$, we consider an initial state
%%%
\begin{align}
         \label{eq:SingleParState}
          \ket{\Psi^{(1)}(0)}&=\ket{0}_1\ket{0}_2\dots\left(\frac{\ket{0}_i+\ket{1}_i}{\sqrt{2}}\right)\dots\ket{0}_{L} \nonumber \\& 
          =\frac{1}{\sqrt{2}}(\ket{\boldsymbol{0}}+\ket{\boldsymbol{1}_i})     
\end{align}
%%%
that is a superposition of vacuum $\ket{\boldsymbol{0}}=\ket{0,0,\ldots,0}$ and states in the single-particle manifold $\ket{l}=\ket{\boldsymbol{1}_i}=\ket{0,0,\ldots,1_i,\ldots,0}$. These kind of initial states can be generated by applying a $\pi/2$ pulse at a given site $i$ of the array and its  stroboscopic time evolution reads
\begin{align}
          \label{eq:EvolvedState}
                  \ket{\Psi^{(1)}(nT)}&=\frac{1}{\sqrt{2}}\left(\ket{\boldsymbol{0}}+\sum_{\lambda}C^{(1)}_{\lambda}e^{-\frac{\mathrm{i}}{\hbar}\varepsilon^{(1)}_{\lambda}nT}\ket{\Phi^{(1)}_{\lambda}(nT)} \right)\ ,  
\end{align}
 where $\ket{\Phi^{(1)}_{\lambda}(t)}$ and $\varepsilon^{(1)}_{\lambda}$ are the Floquet states and quasienergies within the single-photon manifold, respectively.  Note that we have used the notation $\ket{l}=\ket{\boldsymbol{1}_i}$ for simplicity to emphasize that we are working in the single-particle manifold and the initial excitation is located at the $i$-th site.
 
We also require an observable that couples the vacuum and the $N$-photon manifolds. In the single particle case, it is enough to define a set of local quadratures  $\hat{X}_i=(\hat{a}^{\dagger}_i+\hat{a}_i)/\sqrt{2}$ and $\hat{P}_i=\mathrm{i}(\hat{a}^{\dagger}_i-\hat{a}_i)/\sqrt{2}$ of the resonator at site $i$.

Similarly, one can extend all the previous discussion to investigate the two particle manifold $(N=2)$. In the latter case, the initial state should be a superposition of vacuum, states in the single-particle manifold $\ket{\boldsymbol{1}_i}$, $\ket{\boldsymbol{1}_j}$ and states in the two particle manifold $\ket{l}=\ket{\boldsymbol{1}_i,\boldsymbol{1}_j}=\ket{0,\ldots,0,1_i,0,\ldots,1_j,\ldots,0}$ as in Ref.\cite{roushan17}

%%%
\begin{align}
         \label{eq:TwoParState}
          \ket{\Psi^{(2)}(0)}&=\ket{0}_1\ket{0}_2\dots\left(\frac{\ket{0}_i+\ket{1}_i}{\sqrt{2}}\right)\dots\left(\frac{\ket{0}_j+\ket{1}_j}{\sqrt{2}}\right)\ldots\ket{0}_{L} \nonumber \\& 
          =\frac{1}{2}(\ket{\boldsymbol{0}}+\ket{\boldsymbol{1}_i,\boldsymbol{1}_j})+\frac{1}{\sqrt{2}}(\ket{\boldsymbol{1}_i}+\ket{\boldsymbol{1}_j})
          \ .      
\end{align}
%%%
One can generate these initial states  applying two $\pi/2$ pulses at the sites $i$ and $j$ of the array.
Again, here we use the notation $\ket{l}=\ket{\boldsymbol{1}_i,\boldsymbol{1}_j}$  to denote states in the two-particle manifold such that the initial excitation is located at the $i$-th and $j$-th sites.
 To perform spectroscopy in the two-photon manifold, we need to measure observables such as $\hat{X}_i\hat{X}_j$, $\hat{P}_i\hat{P}_j$ and $\hat{P}_i\hat{X}_j$, which gives one information about the spatial correlations. For the detailed explanation of this, we refer the reader to the supplementary material of Ref.\cite{roushan17}.

Depending on the manifold we are interested in, one can choose an observable. For example, $\hat{A}^i=\hat{X}_i$ within the single-particle manifold. In this case the expectation value of the quadrature $\hat{X}_i$ in the state $\ket{\Psi^{(1)}(nT)}$ reads
\begin{equation}
\label{eq:QuadExpVal}
       A^i_n=\ew{\hat{A}_i(nT)}=\ew{\hat{X}_i(nT)}=\sum_{\lambda}|C^{(1)}_{\lambda}|^2\cos(\varepsilon^{(1)}_{\lambda}nT)
       \ .
\end{equation}

One can extend this discussion to the general case of $N$ particles. In that case, one needs to measure higher order correlations. In the general case, after experimentally recording the sequence of stroboscopic measurements  $\{A^l_0,A^l_1,\ldots,A^l_Q\}$ for an initial configuration $\ket{l}$, one can define the discrete Fourier transform

\begin{equation}
          \label{eq:DFT}
                  \tilde{A}^{l}_k=\frac{1}{M}\sum^{Q-1}_{k=0}e^{i\frac{2\pi k}{N} n}A^{l}_n.
\end{equation}

If we let the system evolve for a time $T_{\text{Evol}}$ smaller than the coherence time $T_2$ of the device, we can extract the quasienergies from the peaks of the power spectrum $ |\tilde{A}^l_k|^2$ as a function of $k$. To do so, one should average the power spectrum over all the possible configurations, as follows~\cite{roushan17}

\begin{equation}
          \label{eq:AvePowerSpectrum}
                  \overline{|A_k|^2}=\frac{1}{D_N}\sum^{D_N}_{l=1}|\tilde{A}^l_k|^2.
\end{equation}

 In any case, the location of the peaks in the averaged power spectrum $\overline{|A_k|^2}$ give us information about the quasienergies $\varepsilon^{(N)}_{\lambda}$ and the height of the peaks provides us with the probability amplitude $|C^{(N)}_{\lambda}|^2$, as one can see from Eq.\eqref{eq:QuadExpVal} for $N=1$. Let us assume for example, that we are investigating the model within the $N$-particle manifold. The spectroscopy method can also help us to study how extended the states are in a given basis. For example, for the states like~\eqref{eq:EvolvedState} and~\eqref{eq:TwoParState}, we can define the participation ratio $P=\left(\sum_{\lambda}|C^{(N)}_{\lambda}|^4\right)^{-1}$, where $|C^{(N)}_{\lambda}|^2$ can be extracted from height of the peaks in the power spectrum $ |\tilde{A}^l_k|^2$.

A possible implementation of the Hamiltonian Eq.[1] in the main text could be realized by using  superconducting qubit arrays. In this setup it is possible to achieve couplings of the order of $g_0/2\pi
\sim 50$~MHz with a driving amplitude $g_0\sim g_1$, the onsite energies  can be tuned in the range $0<h_j/2\pi<250$~MHz, and the interaction strength can be $U/2\pi\sim 3.5 g_0= 175$~MHz. In the case of an array with $L=12$ superconducting qubits, one needs to drive the system with a frequency $\omega/2\pi=2.96 g_0/2\pi=148$~MHz. For that driving frequency, the period of the drive will be $T=6.8$ns As the coherence times of devices used in recent experiments such as in Ref.~\cite{roushan17} are of the order of $T_2=5\mu s$, one might be able to observe around $Q=700$ periods of evolution under the drive. To be more concrete, in recent experiments with two photons in nine superconducting qubits, one can study time evolutions up to a time of $T_{\text{Evol}}=250 ns$, which would allow to observe the dynamics up to $Q=36$ periods of the drive.

\end{document}